\documentclass[aps,prl,twocolumn,floats,showpacs]{revtex4}

\usepackage{epsfig}
\usepackage{epstopdf}
\usepackage{graphicx}
\usepackage{amsmath}
\usepackage{amssymb}
\usepackage{units}
\usepackage[usenames,dvipsnames]{color}

\newcommand{\beq}{\begin{equation}}
\newcommand{\eeq}{\end{equation}}
\newcommand{\bea}{\begin{eqnarray}}
\newcommand{\eea}{\end{eqnarray}}

\begin{document}
\title{Advanced experimental applications for x-ray transmission gratings Spectroscopy using a novel grating fabrication method}
\date{\today}
\author{G.~Hurvitz$^{1}$, Y.~Ehrlich$^{1}$, G.~Strum$^{2}$, Z.~Shpilman$^{1}$, I.~Levy$^{1}$ and M. Fraenkel$^{1}$}
\affiliation{Plasma Physics Department$^{1}$, and Solid State Department$^{2}$, Applied Physics Division, Soreq NRC, Yavne, Israel\\}
%
\begin{abstract}{
A novel fabrication method for soft x-ray transmission grating and other optical elements is presented. The method uses Focused-Ion-Beam (FIB) technology to fabricate high-quality free standing grating bars on Transmission Electron Microscopy grids (TEM-grid). High quality transmission gratings are obtained with superb accuracy and versatility. Using these gratings and back-illuminated CCD camera, absolutely calibrated x-ray spectra can be acquired for soft x-ray source diagnostics in the 100-3000 eV spectral range. Double grating combinations of identical or different parameters are easily fabricated, allowing advanced one-shot application of transmission grating spectroscopy. These applications include spectroscopy with different spectral resolutions, bandwidths, dynamic ranges, and may serve for identification of high-order contribution, and spectral calibrations of various x-ray optical elements.}
\end{abstract}
\maketitle
\section{I. Introduction}
Spectroscopy of both laboratory and astrophysical x-ray sources require dispersive elements with adequate spectral and spatial resolutions. These elements should offer the ability to acquire quantitative information on spectral composition of the emitted radiation. A widely used dispersive elements is the Transmission Grating (TG), which, when combined with an adequate x-ray detector, consists the Transmission Grating Spectrometer (TGS)\cite{1,1.5}.\\
TGS characteristics should be application tailored. For example, for the soft x-ray region (photon energy of few tens of eV to several keV), as in common laser produced and Z-pinch plasmas, the TG should include free standing high-Z material bars with grating period in the order of $1~\mu$. Production of such gratings is challenging, especially when high bar density ($>1000$ lp/mm), high bar length over width ratio ($>10$), and high accuracy structures are required.\\
Traditional TG fabrication methods include electron-beam lithography\cite{2} and mechanical ruling and replication\cite{3}, resulting in uncertainty and non-uniformity of grating bars width, thickness, and period. These effects increase as the grating period becomes smaller, limiting the ability to use TG  theoretical transmission calculations as a basis for absolute calibration of acquired spectra.\\*
In this paper we introduce a novel cost-effective method for transmission grating fabrication using TEM grid as substrate, and Focused Ion Beam (FIB) machine for "writing" the grating. This robust fabrication process ensures production of high accuracy and quality TGs. In addition, this method allows high flexibility in grating design, including double grating combination. These designs enable new experimental applications, some of which are described in section V.
\section{II. TRANSMISSION GRATING SPECTROMETER CONCEPT}
A TGS is a simple experimental apparatus, containing a radiation source, a TG, and a detector. Common detectors include x-ray sensitive films and CCD cameras. These detectors have been calibrated for the soft x-ray regime, and their spectral response is used with relatively high reliability\cite{4,4.1,4.2}.\\
The TGS performance is determined by several TG characteristics, including TG bars material and thickness, grating period, duty cycle (spacing between bars divided by grating period), and structure accuracy. For example, the material and bars thickness should be chosen so that transmission through the grating bars would be negligible throughout the entire measured spectrum. Gold bars with thickness of several hundred nm are commonly used, as gold has high absorption in the soft x¬-ray regime and is also suitable for common fabrication techniques in lithography and microelectronics.\\
The grating dispersion angle is given by the Bragg equation\cite{5}:
\begin{equation}
sin(\theta)=\frac{m\lambda}{d} \label{eq-1}
\end{equation}
where $\theta$ is the diffraction angle, $\lambda$ is the photon wavelength, $m$ is the order of diffraction, and $d$ is the grating period. Due to fabrication limitations, grating periods much smaller than $1~\mu$ are rare, and thus the resulting diffraction angle is usually in the order of a degree. The grating active area governs the resolving power, which is limited by:
\begin{equation}
\frac{\lambda}{\Delta\lambda}<mN \label{eq-2}
\end{equation}
where $\Delta\lambda$ is the wavelength resolution and $N$ is the total number of grating periods used for the diffraction. The transmission efficiency of the TG is derived from Fraunhofer diffraction theory\cite{5,6}:
%
%
\begin{align}
    &\eta_m=(\frac{sin(m\pi(a/d))}{m\pi})^2(1+c_1^2-2c_1c_2), ~~~m\ne0  \label{eq-3}\\
    &\eta_0=(\frac{a}{d})^2+(1-\frac{a}{d}^2)c_1^2 + 2\frac{a}{d}(1-\frac{a}{d})c_1c_2  \label{eq-4}
\end{align}
where $a$ is the grating's inter-bar spacing, $a/d$ is the grating duty cycle, $c_1=exp(-2\pi n_2z_0/\lambda)$
and  $c_2=cos(2\pi (n_1-1)z_0/\lambda)$ contain the wavelength dependence through the grating's material
nonlinear index of refraction $n=n_1+in_2$, and the bars thickness $z_0$.
 These parameters should only be taken into consideration where the grating bars are partially transmissive, otherwise $c_1$ approaches zero. In this common case and $a/d=0.5$, diffraction efficiency into each of the first orders is $1/\pi^2$, and high orders efficiency decreases as $1/m^2$. The grating duty cycle ($a/d$) governs the energy distribution into each diffraction order, and specifically, a duty cycle of $a/d=1/q$, eliminates diffraction into orders divisible by $q$. Thus, a symmetrical ($a/d=0.5$) grating has no even diffraction orders.\\
Diffraction efficiency calibration of a specific TG for the x-ray regime is a cumbersome task that is usually performed using an absolutely calibrated wide-band soft x-ray source, such as a synchrotron beam-line\cite{6,7}. Even when calibration is performed for specific x-ray energies, calculated data is used to complete the efficiency curve for the desired spectral range\cite{6,7}. TG efficiency can be calculated by eq.\ref{eq-3} with reasonable accuracy under the assumption of high quality grating: perfect periodical structure and bars of sufficient thickness with perfect rectangular cross-section. Imperfections in grating fabrication lead to spectral uncertainties and disagreement between calculated and practical efficiencies. Unequal spacing between grating bars scatters energy between diffraction orders, asymmetrical cross-sections of grating bars lead to asymmetry between positive and negative orders, and non-rectangular bars might enhance transmission at certain photon energies\cite{6}. Numerical methods can be applied\cite{8} to estimate the influence of these imperfections on the TG transmission efficiency, with limited accuracy. Therefore, the calculated transmission efficiency can only be assumed when high quality and high accuracy gratings are used. Consequently, grating fabrication technique capabilities are a key factor for obtaining absolutely calibrated spectra of soft x-ray source with high accuracy.
\section{III. FIB-BASED TG FABRICATION METHOD}
\begin{figure}
\vspace{-0.0cm}
\centering
\includegraphics[width=0.9\columnwidth]{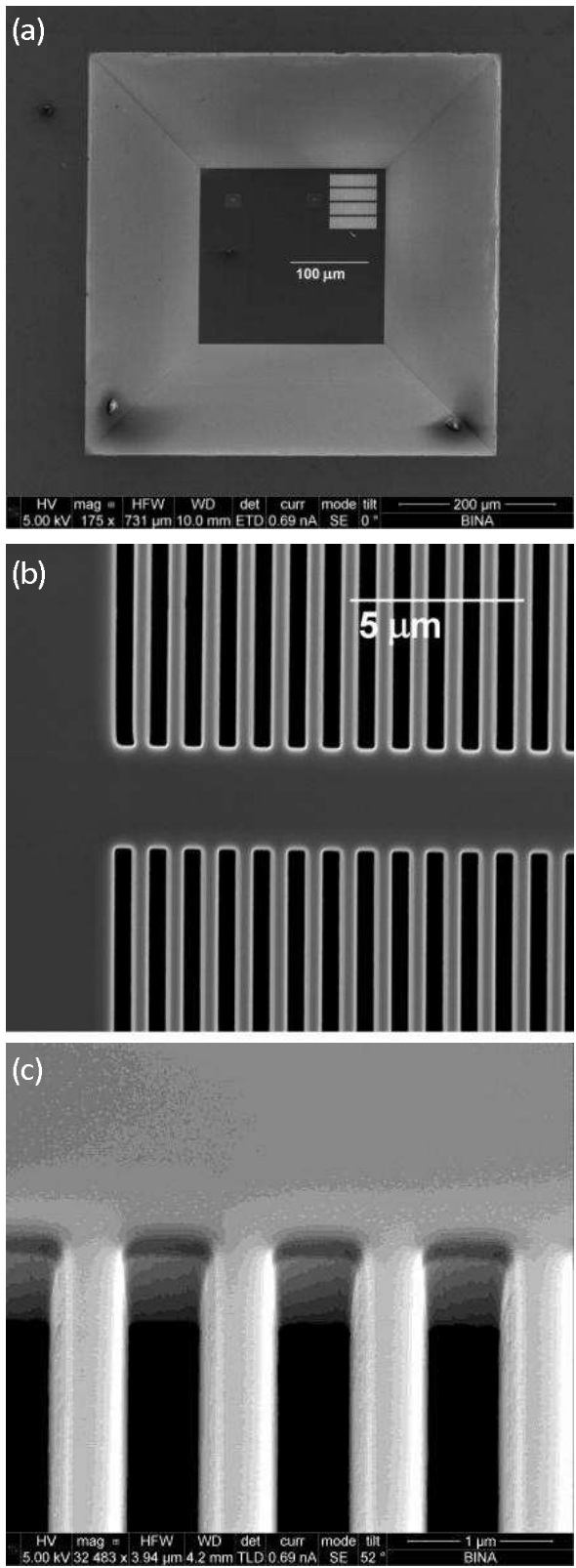}
\vspace{-2.0cm}
     \caption{HR-SEM picture of TEM$\times$FIB fabricated TG with period of $1~\mu$ and duty cycle of 50\%. (a) The whole TEM-grid, (b) a closer view, the exact periodicity can be seen, (c) close up view showing the 3D quality of the FIB milling: Upper layer is the SiN membrane, followed by Ti layer and below is brighter Au layer.}
  \vskip -.1truein
  \label{fig-1}
\end{figure}

The presented novel fabrication method uses transmission electron microscope (TEM) grid as the grating's substrate, and the inter-bars spacing is "written" using focused ion beam (FIB).\\
The grating substrate is prepared using commercial TEM grids. These grids are composed of thin SiO$_2$ or Si$_3$N$_4$ membrane held by a thick Si wafer frame. A $50$ nm thick membrane over an area of $500 \times 1500 ~ \mu m ^2$ is a standard TEM shelf item.
The Si frame is used as a mechanical holder for the grating as part of the TGS.
The membrane is initially coated with a thin Ti adhesion layer, followed by evaporation or sputtering of the desired Au layer ($\sim 5000~\AA$), resulting with an excellent cost-effective substrate for TG.
For dense TG ($>1000$ mm), a thin substrate should be used, and an annealing process may be required in order to eliminate any residual stress in the composed membrane.\\
FIB\cite{9}, which is used commercially for more than two decades, uses Gallium ion (Ga$^+$) beam accelerated up to $30$ keV and focused down to $5$ nm resolution. These energetic heavy ion beams are used mainly for milling, by sputtering substrate material. Modern FIB systems comprise high resolution scanning electron microscope (HR-SEM), which enables in-situ sample examination. This allows process refining by calibration of production parameters, enabling the achievement of accurate high quality products with just few iterations.\\
FIB was used to mill TG patters into the TEM-grid based substrate. A software procedure specifies the desired TG layout, and controls its fabrication by moving the substrate to the appropriate locations. In order to achieve satisfying results (especially in term of precise control of duty cycle), it is necessary to tune FIB parameters, such as: beam focal size, beam voltage and current, and writing speed. This stage can usually be completed during milling of few TG bars, while monitoring the production using the HR-SEM. After the FIB parameters are set, the desired TG layout is uploaded to the FIB software and the process is automatically performed by the machine. An example of the TEM-FIB (TEM$\times$FIB) fabricated TG with $1000$ lp/mm and $a/d=0.5$ is shown in Fig. \ref{fig-1}.
The grating high quality in terms of period, equal bar width, thickness and spacing is clearly seen over the entire grating area.\\
A simulation code based on the Huygens integral was used to plan TG schemes and check sensitivity to various parameters.
Contrary to eqs. \ref{eq-3}-\ref{eq-4}, the code simulates real TG structures, such as finite grating and contribution of the support beams.
\section{IV. X-RAY DISPERSION EXPERIMENTS USING FIB FABRICATED TG}
\begin{figure}
\centering
\vspace{-0.0cm}
\includegraphics[width=1\columnwidth]{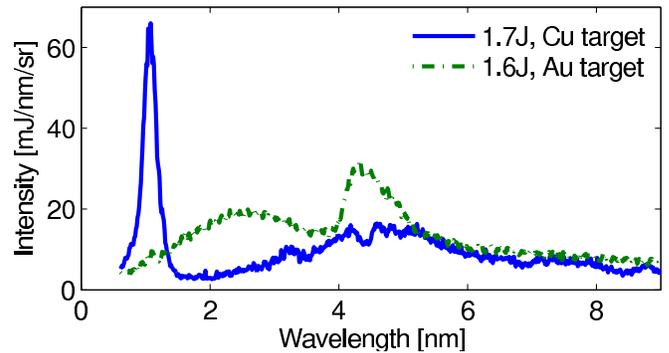}
\vspace{-0.6cm}
\caption{Absolutely calibrated Au and Cu spectra as acquired by the TGS. $1$ ns Nd:YAG laser pulse was focused on a disk target. Absolute calibration was performed using calculated TG transmission efficiency, geometrical factors and published CCD detector response.}
\label{fig-2}
\end{figure}

The gratings were used for x-ray spectroscopy experiments using laser produced plasma as the diagnosed x-ray source. An Nd:YAG laser pulse of $1$ ns duration and energy of up to $20$ J per pulse at SHG ($532$ nm) was focused on a metal foil located inside a vacuum vessel, to create high intensity x-ray plasma source. The laser is focused using an aspheric lens with a focal length of $450$ mm to a spot of about $50~\mu$ diameter on target. The x-ray source size was measured using an x-ray pin-hole camera coupled to a CCD. The TGS comprised the TG, located $450$ mm from the source, and a back-illuminated x-ray sensitive CCD camera (Andor DX434) located $1$ m from the grating, which was used as the spectrometer detector. Examples of 
absolutely calibrated Au and Cu spectra as acquired by the TGS are shown in Fig.\ref{fig-2}. The calibration is calculated using TG spectral sensitivity\cite{10}, CCD quantum efficiency\cite{7,11} and geometrical factors derived from the specific experimental scheme.
\section{V. ADVANCED EXPERIMENTAL APPLICATIONS USING DOUBLE GRATING APPARATUS}
\begin{figure}
\centering
\vspace{-0.0cm}
\includegraphics[width=1\columnwidth]{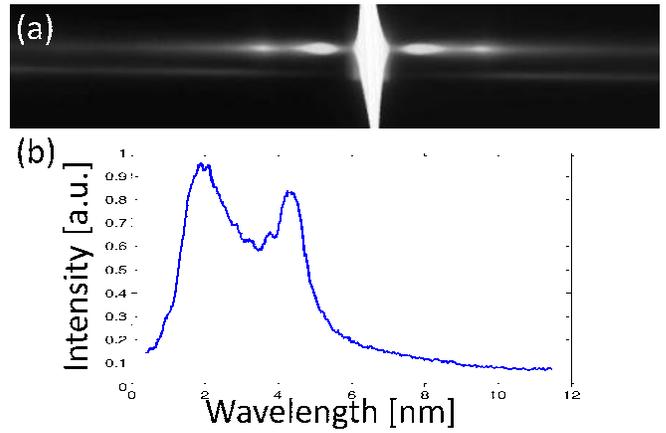}
\vspace{-0.6cm}
\caption{Double-grating combination. 
(a) Au spectra acquired in a single-shot experiment by this device. Upper spectrum has a wide spectral range reaching $\sim 14$ nm, with lower resolution. Lower spectrum has a narrower spectral range without saturation near $2$ nm, and with enhanced resolution. 
(b) Combined spectrum derived from both TGs gives a wide spectral range with a large dynamic range and improved resolution, taken in a single-shot.}
\label{fig-4}
\end{figure}

In addition to the capabilities in standard grating fabrication, our novel TEM$\times$FIB processing procedure allows the fabrication of composed schemes, as double gratings of identical or different parameters, on a single TEM-grid. Fig. \ref{fig-3} shows an example of spectra acquired by double-grating combination of identical gratings, showing superb spectra reproducibility. Thus, these double-gratings schemes may be used for several advanced TGS applications:
\begin{figure}
\centering
\vspace{-0.0cm}
\includegraphics[width=1\columnwidth]{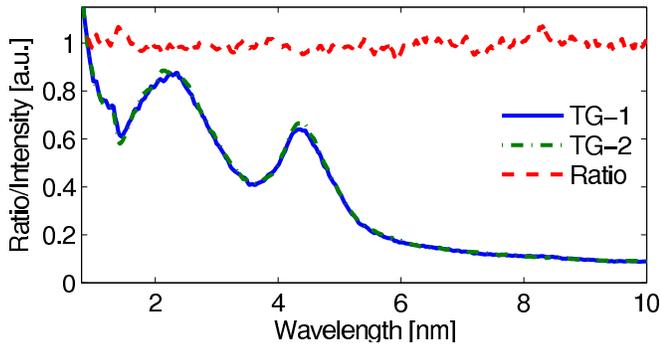}
\vspace{-0.5cm}
\caption{Two spectra acquired by a typical double-grating combination of identical gratings.
Superb spectra reproducibility points to the high quality production capabilities.}
\label{fig-3}
\end{figure}
\subsection{1. One-shot spectroscopy with different dispersion power (Fig. \ref{fig-4})}
\begin{figure}
\centering
\vspace{-0.5cm}
\includegraphics[width=1\columnwidth]{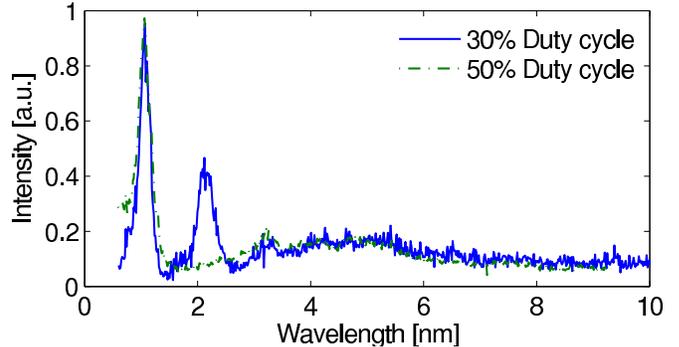}
\vspace{-0.5cm}
\caption{Single-shot experiment with Cu target. The spectrum contains a strong L shell peak at $1.1$nm and a broad thermal radiation at $\sim 5$ nm. The two spectra were taken by single device having two TGs with identical periods but different duty cycles of $50\%$ and $30\%$. The $30\%$ duty cycle spectrum contains a peak at $2.2$ nm ($2^{nd}$ order of the L shell), while the $50\%$ duty cycle spectrum is free of even diffraction orders. Higher orders of the peak may also be recognized.}
\label{fig-5}
\end{figure}

One TEM-grid having two gratings with periods of $2~\mu$ and $0.6~\mu$ is used. In a one-shot experiment, a broad x-ray spectrum is acquired with lower spectral resolution using the first grating, while a partial band is acquired by the second grating with higher dispersion power. This allows better spectral resolution for a specific spectral band, and also allows resolving (for instance) high-energy spectrum that is hidden inside the zero-order of the low-resolution spectrum.\\*
This same example also demonstrates enhancement of the dynamic range using double measurement with different grating's efficiencies.
Since the energy spread on the detector is
determined by the grating's period,
the higher the spectral resolution, the lower the flux on the detector is.
The total transmitted intensity depends on grating's active area and duty cycle, therefore an appropriate choice of period, duty cycle and dimensions of each grating may dramatically expand the dynamic range.
\subsection{2. One-shot spectroscopy with different $a/d$ ratio (see Fig. \ref{fig-5})}
\begin{figure}
\centering
\vspace{-0.0cm}
\includegraphics[width=1\columnwidth]{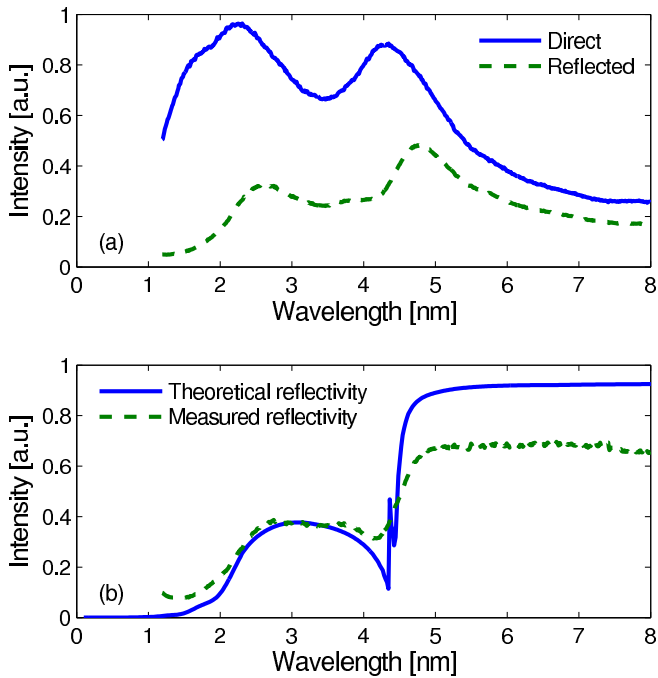}
\vspace{-0.5cm}
\caption{One-shot spectral calibration of x-ray grazing incidence mirror. (a) two spectra diffracted by identical gratings (produced on a single TEM-grid), one is reflected by a carbon mirror at an angle of $3.3^0$, while the other is a reference. (b) spectral mirror reflectivity: an experimental reflectivity extracted from the upper graph, a calculated reflectivity given by Henke et. al.\cite{10}.}
\label{fig-6}
\end{figure}

Use of gratings with high-accuracy duty cycle is a pre-condition for reliable identification and ellimination of high orders contribution to the spectrum.
TG spectroscopy with different $a/d$ ratio allows the suppression of different high diffraction orders (see eq. \ref{eq-3}). One-shot acquisition with several $a/d$ ratios allows identification of different high-order contributions to the spectrum, which can be subtracted in post-processing. 
\subsection{3. One-shot spectral calibration of x-ray optics (see Fig. \ref{fig-6})}
Although spectral response of x-ray optical elements as filters and grazing-incidence mirrors can be calculated with relatively high reliability \cite{10}, measurement of the spectral response of a specific optical element can reveal deviations from the calculated performance. This is mainly due to uncertainties in manufacturing parameters (filter thickness, mirror roughness) and gradual degradation of the element, due to oxidation, adhesion of impurities, etc.
A simple experimental scheme for one-shot measurement of the spectral response is presented (filter transmission, mirror reflection, etc.).
In this scheme, one TEM-grid having two identical gratings is used in combination with an optical element such as an x-ray mirror or a filter which has to be characterized.
The setup allows spectrum from the first grating to arrive directly to the detector, to serve as the reference spectrum.
The spectrum from the second grating is transmitted through a filter or reflected by a mirror in grazing incidence angle,
and may be acquired on the same detector.
The spectral performance function of the optical element is derived from the ratio of two spectra intensities in a single-shot. As an example, Fig. \ref{fig-6} shows a reference spectrum and the reflected spectrum from a carbon mirror in grazing incidence ($3.3^0$). The ratio is compared to the theoretical reflection curve of the mirror\cite{10}.
\section{VI. Conclusions}
The TEM$\times$FIB method presented here allows the fabrication of high-quality and high-accuracy TGs in a robust process with scalability to larger areas, control of all grating's parameters, flexibility to parameters change, reproducibility, and relatively low-cost per item. 
Advanced experimental possibilities using TGS are easily available by few gratings combinations with identical or different parameters. This method can also be used to fabricate other x-ray optical elements, as pin-hole sets of different diameters, and gratings for cold atom experiments. Moreover, practically any design is possible with the suggested technique, with great accuracy and low effort. This ability opens-up opportunities for high-accuracy x-ray source diagnostics using TGS and pin-hole camera schemes.
%
%

\end{document}